\documentclass[11pt]{article}
\pdfoutput=1
\usepackage{jheppub}
\usepackage{graphicx}
\usepackage{amssymb}
\usepackage{amsmath,amssymb}
\usepackage{slashed}
\usepackage{hyperref}
\usepackage{caption}
\usepackage{xcolor}
\usepackage{dsfont}
\usepackage{verbatim}
\usepackage{subfig}

\newcommand\beq{\begin{equation}}
\newcommand\eeq{\end{equation}}
\newcommand\be{\begin{equation}}
\newcommand\ee{\end{equation}}

\newcommand\ra{\rightarrow}

\title{Spontaneously Broken Subsystem Symmetries}

\preprint{UTTG-21-2021}

\author{Jacques Distler}
\author{Andreas Karch}
\author{Amir Raz}

\affiliation{University of Texas, Austin, Physics Department, Austin TX 78712, USA}

\emailAdd{distler@golem.ph.utexas.edu,karcha@utexas.edu,araz@utexas.edu}

\abstract{We investigate the spontaneous breaking of subsystem symmetries directly in the context of continuum field theories by calculating the correlation function of charged operators. Our methods confirm the lack of spontaneous symmetry breaking in some of the existing continuum field theories with subsystem symmetries, as had previously been established based on a careful analysis of the spectrum. We present some novel continuum field theory constructions that do exhibit spontaneous symmetry breaking whenever allowed by general principles. These interesting patterns of symmetry breaking occur despite the fact that all the theories we study are non-interacting.}

\begin{document}
\maketitle

\section{Introduction}

Fracton lattice models \cite{paramekanti2002ring,Chamon:2004lew,haah2011local,Nandkishore_2019,Pretko_2020} have forced a rethinking of some of the longest held believes about quantum field theories. Standard lore would posit that, at very long distances, any quantum field theory should either flow to a gapped system, potentially described by a non-trivial topological field theory (TFT), or a gapless system described by a conformal field theory (CFT). Fractonic lattice models do not fit into either of these frameworks; their limited mobility excitations as well as their system size dependent groundstate degeneracy is not describably by any standard TFT or CFT.

A genuine continuum field theory framework capable of describing fractonic physics has been laid out in \cite{Seiberg_2020,Seiberg:2020bhn,Seiberg:2020wsg}. The major new ingredient in these field theories are subsystem symmetries: global symmetries that do not act on the entire system but only on degrees of freedom localized along certain sub-manifolds. Such subsystem symmetries are often present in fractonic lattice models; for the exotic field theories constructed in \cite{Seiberg_2020,Seiberg:2020bhn,Seiberg:2020wsg} they become a definining feature. While some of these new field theories don't even exhibit genuine fractonic excitations, they all have some kind of subsystem symmetry.

At face value these exotic field theories look somewhat conventional. The simplest examples are free field theories; in fact for many examples the symmetries prevent any relevant or marginal interactions. On the surface the only new features are unusual kinetic terms. For example, in the simplest 2+1 dimensional model of  \cite{Seiberg:2020bhn} based on a single real scalar, the standard $(\partial_x \phi)^2 + (\partial_y \phi)^2$ kinetic term gets replaced with $(\partial_x \partial_y \phi)^2$. This is the lowest dimension kinetic term consistent with a subsystem shift symmetry demanding the action be invariant under shifts of $\phi$ that can depend on either $x$ or $y$, $\delta \phi = f_x(x) + f_y(y)$. This kinetic term, however, leads to dramatic consequences. Field configurations with arbitrary large discontinuities in the $x$ direction have low energy, and are hence unsuppressed in the path integral, as long as they are smooth in $y$, and vice versa. The presence of these discontinuous field configurations gives rise to a surprising interplay between UV and IR physics which was originally observed in \cite{Seiberg:2020bhn,Seiberg:2020wsg}, and later expanded on in \cite{you2021fractonic,you2021fractonic2,gorantla2021lowenergy}. Most notably, many quantities in the continuum theory remain sensitive to the lattice spacing of the underlying discrete model, that is to UV scale physics.

One place where this UV/IR mixing becomes important is when trying to understand whether the subsystem symmetries are spontaneously broken. Already in the original examples of \cite{Seiberg_2020,Seiberg:2020bhn,Seiberg:2020wsg} the issue of symmetry breaking turned out to be quite subtle. The theory has a large number of classical zero energy modes, carrying a large momentum in one direction but no momentum in the other. One could identify these modes with Goldstone modes of a putatively broken subsystem symmetry. However, quantum mechanically these modes get pushed to energies of order the UV cutoff. The symmetry is in fact restored. Closely related is the fact that the lightest states charged under the subsystem symmetry that could detect any breaking get pushed to UV scale energies as well.

A very good probe of symmetry breaking are correlation functions of charged operators. One thing we do in this work is to calculate these correlation functions directly in the continuum theories of \cite{Seiberg_2020,Seiberg:2020bhn,Seiberg:2020wsg} and verify that the subsystem symmetries do, in fact, remain unbroken as suggested by the spectrum. This same conclusion has also been recently obtained in \cite{gorantla2021lowenergy}, where the same correlation functions were obtained by carefully taking the limit of the discretized theory. 

Given these results, one may wonder whether and when spontaneous breaking of subsystem symmetries in a continuum field theory can happen. As in ordinary quantum field theories, where the Mermin-Wagner-Coleman theorem forbids breaking of a continuous symmetry in 1+1 dimensions and below, spontaneous symmetry breaking should not always be allowed. A theorem due to Batista and Nussinov \cite{Batista:2004sc} shows that the criteria of whether a subsystem symmetry acting on a $n$ dimensional submanifold can be broken or not are the same as for an ordinary symmetry for a field theory living in an $n$ dimensional spacetime. For sufficiently high $n$ there is nothing that forbids breaking and so one may wonder how to realize this phenomenon in exotic field theories.

One option to analyze the spontaneous breaking of subsystem symmetries is to work directly with the lattice models. Indeed, one of the early studies of fractons showed that the spontaneous breaking of a discrete subsystem symmetry accounts for the large ground state degeneracy in the X-cube model \cite{Vijay_2016}. More recently progress has been made in studying the phase transition between the broken and unbroken phase of discrete subsystem symmetries in lattice models \cite{you2021fractonic2,PhysRevB.98.035112,You_2020,Qi_2021,lake2021subdimensional,Bulmash_2018}. The breaking of the continuous subsystem symmetry of \cite{Seiberg:2020bhn,Seiberg:2020wsg} can also be achieved by taking a different limit of the lattice model, namely an infinte volume limit at finite lattice spacing \cite{gorantla2021lowenergy}. While this limit does give rise to a broken phase, it does not seem to be amendable to continuum field theory language. 

In contrast to these lattice constructions, in this work we present some novel continuum field theories that do exhibit subsystem symmetry breaking. Interestingly enough, all these theories are free theories, just like the original work of
\cite{Seiberg_2020,Seiberg:2020bhn,Seiberg:2020wsg}, and so our results are exact. Additionally, we find that whenever the theorem of \cite{Batista:2004sc} does allow symmetry breaking, our models indeed do lead to spontaneously broken subsystem symmetries.

The outline of this work is as follows. In section 2 we will review why we chose correlations of charged operators to detect symmetry breaking and show that our methods reproduce the known results for the theories of \cite{Seiberg_2020,Seiberg:2020bhn,Seiberg:2020wsg}, consistent with the recent paper \cite{gorantla2021lowenergy}. In section 3 we introduce the classical XY-plaquette model as one example of a field theory that does in fact lead to a spontaneously broken symmetry in dimensions where this is not prohibited by the theorem of \cite{Batista:2004sc}. In section 4 we generalize our model to a large class of closely related theories that all follow the same pattern. We end with a discussion and future directions in section 5.

\section{Diagnosing the spontaneous breaking of subsystem symmetries}

Let us consider a field theory living in $d$ spacetime dimensions  with a subsystem symmetry acting on $n$ dimensional submanifolds. According to the modified Mermin-Wagner-Coleman theorem \cite{Batista:2004sc}, the rules for whether spontaneous breaking of these subsystem symmetries can or can not occur are equivalent to the standard considerations for a quantum field theory living in $n$ dimensions. In particular, spontaneous breaking of a continuous global subsystem symmetry can only happen if $n \geq 2$. However this theorem in and of itself does little to predict if a specific subsystem symmetry is spontaneously broken, or even how to diagnose if such a breaking occurs. 

The classical approach to diagnosing a spontaneously broken symmetry is to define a local order parameter that is charged under the symmetry, and then calculate if this order parameter acquires an expectation value in the vacuum state. This computation is delicate as the expectation value of a charged operator vanishes for any finite size system which obeys the symmetry. Hence the correct procedure is to calculate the expectation value of the order parameter in the presence of a symmetry breaking background field, which is then taken to zero. \footnote{For a modern mathematical description of spontaneous symmetry breaking see \cite{strocchi2005symmetry}.}

The canonical example of this procedure is to consider a complex field in $d$ dimensions, $\phi$, and a Lagrangian with a Mexican hat potential
\begin{equation}
    S = \int d^d x \left[ \partial_\mu \phi^\dagger \partial^\mu \phi 
    + g \left( \phi^\dagger \phi - v^2 \right)^2\right].
\end{equation}
This theory has a continuous $U(1)$ symmetry which shifts $\phi$ by a phase, so $\phi$ is a natural order parameter for this theory. Notice that the minimum of the potential is not invariant under this symmetry, so classically we would expect that the vacuum expectation of $\phi$ to be $|\left<\phi \right>| = v$. However this may not be the case in the quantum field theory. To compute the true expectation value of $\phi$ we need to introduce a symmetry breaking term to the action
\begin{equation}
    \delta S_h = \int d^d x ~h \phi, 
\end{equation}
and then compute the expectation value of $\phi$ in the presence of this term. The true expectation value of $\phi$ will be given by the limit
\begin{equation}
    \left< \phi \right> = \lim_{h \ra 0} \left< \phi \right>_h = 
     \lim_{h \ra 0} \int D\phi D\phi^\dagger ~\phi ~e^{-S - \delta S_h} .
\end{equation}

Though this approach was used to prove the Mermin-Wagner-Coleman theorem for subsystem symmetries in a lattice theory \cite{Batista:2004sc}, it not clear how to define an order parameter for a subsystem symmetry in a continuum field theory, or how to introduce a symmetry breaking term that only breaks the desired subsystem symmetry.

A different approach to diagnosing spontaneous symmetry breaking is to compute the long range correlation functions of charged operators, and see if they vanish. This approach is used in the context of higher form symmetries to diagnose if a $p$-form symmetry is spontaneously broken \cite{Gaiotto_2015,lake2018higher}. For the above example of the complex scalar field $\phi$, one can compute the two point function
\begin{equation}
    \left< \phi^\dagger(0) \phi(r) \right>.
\end{equation}
If the field theory is local then the expectation values of a product of distant operators should factorize, and thus
\begin{equation}
\lim_{r \rightarrow \infty} \left< \phi^\dagger(0) \phi(r) \right>  =     \big< \phi^\dagger\big> \big< \phi\big> = \big< \phi\big>^\dagger \big< \phi\big>.
\end{equation}
Hence if $\lim_{r \rightarrow \infty} \left< \phi^\dagger(0) \phi(r) \right>  = 0 $ then the symmetry is not spontaneously broken, while if $\lim_{r \rightarrow \infty} \left< \phi^\dagger(0) \phi(r) \right>  \neq 0 $ then $\phi$ acquires an expectation value and the symmetry is spontaneously broken. We note that the local charged operator $\phi$ may need to be renormalized to give a finite answer. In such a case the UV renormalization needs to be distinguished from the IR behavior of the 2-point function. 

This same method can be used to identify spontaneous symmetry breaking of subsystem symmetries, though we must ensure that the long distance correlation functions used to diagnose symmetry breaking is uncharged under the symmetry. Thus, for a subsystem symmetry, the correlation functions used to identify symmetry breaking will not be 2-point functions, but rather higher point correlation functions. We present an example of such a correlation function in the following subsection.

\subsection{Symmetry breaking in the XY-plaquette model}

As simple example of identifying spontaneous symmetry breaking of subsystem symmetries, we shall look at continuum field description of the XY plaquette model  \cite{Seiberg:2020bhn,Seiberg:2020wsg,paramekanti2002ring}. In \cite{Seiberg:2020bhn,Seiberg:2020wsg} it was claimed that no spontaneous symmetry breaking happens, even in $3+1$ dimensions where spontaneous symmetry breaking is permitted. We would like to verify this result using our diagnostic for spontaneous symmetry breaking.

The original XY-plaquette model in $d+1$ dimensions has a continuum field theory formulation consisting of a compact scalar field $\phi$, and the action (in Euclidean signature)
\begin{equation}
    L = \frac{\mu_0}{2}\partial_0\phi \partial_0\phi + \frac{1}{2\mu} \sum_{1 \leq i<j \leq d} \partial_i\partial_j \phi\partial_i\partial_j \phi .
\end{equation}
This model has the global subsystem shift symmetry which acts on the primary $1+(d-1)$ dimensional planes and is given by
\begin{equation}
    \phi \ra \phi + \sum_{i=1}^d f_i(x_i),
\end{equation}
where $f_i(x_i)$ is an arbitrary function of the $i$'th coordinate.

To diagnose if this symmetry is spontaneously broken we should compute  a long-range correlator of charged operators, which for this theory is the operator $e^{i\phi}$ (as $\phi$ is not a good operator in the continuum field theory \cite{Seiberg:2020bhn}). Furthermore, we need the correlator we compute to be uncharged under the global subsystem shift symmetry. The simplest correlator that meets this criteria is the equal time correlator of four $e^{\pm i\phi}$ operators inserted on one of the spatial planes
\begin{equation} \label{eq:corr_xy_org}
    \left<e^{i [\phi(0,0,0) -\phi(0,x_1,0)-\phi(0,0,x_2) + \phi(0,x_1,x_2)]} \right> ,
\end{equation}
where without loss of generality we took the operators to lie in $x_1$--$x_2$ plane, and where the remaining spatial coordinates are equal.

As this is a Gaussian theory, we can take the functional integral directly, and we find that
\begin{equation}
    \left<e^{i [\phi(0,0,0) -\phi(0,x_1,0)-\phi(0,0,x_2) + \phi(0,x_1,x_2)]} \right> = e^{- \frac{2}{\mu_0} K_d(x_1,x_2)},
\end{equation}
where
\begin{equation}
\begin{aligned}
    K_d(x,y) &= \int \frac{d\omega}{2\pi} \int \frac{d^d k}{(2\pi)^d} \frac{(1-\cos(k_1x))(1-\cos(k_2y))}{\omega^2 + \frac{1}{\mu \mu_0} \sum_{i\neq j} k_i^2 k_j^2}\\
    &= \frac{ \sqrt{\mu \mu_0}}{2}  \int \frac{d^d k}{(2\pi)^d} \frac{(1-\cos(k_1x))(1-\cos(k_2y))}{ \sqrt{\sum_{i\neq j} k_i^2 k_j^2 }}
\end{aligned}
\end{equation}

Starting with the case $d=2$, we have that
\begin{equation}
    K_2(x,y) = \frac{ \sqrt{\mu \mu_0}}{8 \pi^2} \int d^2 k \frac{(1-\cos(k_1x))(1-\cos(k_2y))}{|k_1 k_2|}.
\end{equation}
This integral is UV divergent, and so must be regulated. As the integral factorizes, it is natural to introduce independent cutoffs in the $x$ and $y$ directions so that this integral becomes \footnote{We note that the form of \eqref{eq:K_2_sepregs} should not depend on the particular regularization scheme. For example on can compute $K_2(x,y)$ using the standard spherical momentum cutoff of $k_1^2+k_2^2 \leq \Lambda^2$ and find that $K_2(x,y) = \frac{\sqrt{\mu \mu_0}}{2 \pi^2} \left(\gamma + \log(x\Lambda) \right)\left(\gamma + \log(y\Lambda) \right) - \frac{\sqrt{\mu \mu_0}}{48}  + O(1/\Lambda)$, which agrees with \eqref{eq:K_2_sepregs} up to an additive constant. }
\begin{equation} \label{eq:K_2_sepregs}
\begin{aligned}
    K_2(x,y) &= \frac{ \sqrt{\mu \mu_0}}{2 \pi^2} \int_{0}^{\Lambda_x} d k_1 \frac{(1-\cos(k_1x))}{k_1} \int_{0}^{\Lambda_y} d k_2 \frac{(1-\cos(k_2y))}{k_2} \\
    & = \frac{\sqrt{\mu \mu_0}}{2 \pi^2} \left(\gamma + \log(x\Lambda_x) + O(1/\Lambda_x) \right)\left(\gamma + \log(y\Lambda_y) + O(1/\Lambda_y) \right) .
\end{aligned}
\end{equation}
 This UV divergence cannot be regulated by a renormalization of the operators (or equivalently by local counter term) due to the presence of the terms proportional to $\log(x)\log(\Lambda_y)$ and  $\log(y)\log(\Lambda_x)$. Not only do these terms make the result unregularizable, their presence indicates a UV-IR mixing in the charged states of this model, as observed in \cite{Seiberg:2020bhn,Seiberg:2020wsg}. This same correlation function was computed in the lattice theory in  \cite{paramekanti2002ring} and \cite{gorantla2021lowenergy}, and both found the same result.\footnote{One can directly compare \eqref{eq:K_2_sepregs} to equation (66) in \cite{paramekanti2002ring} and equation (5.33) in \cite{gorantla2021lowenergy}.}

We can do a similar calculation in $d=3$, where we can take the $k_3$ integral by introducing a cutoff, so
\begin{equation}
\begin{aligned}
    K_3(x,y) &=\frac{\sqrt{\mu \mu_0}}{16 \pi^3}  \int d^3 k \frac{(1-\cos(k_1x))(1-\cos(k_2y))}{\sqrt{k_1^2 k_2^2 + k_3^2(k_1^2 + k_2^2)}} \\
    &= \frac{\sqrt{\mu \mu_0}}{16 \pi^3}  \int d^2 k \frac{(1-\cos(k_1x))(1-\cos(k_2y))}{\sqrt{k_1^1 + k_2^2}} \left(2\log(\Lambda_z) + \log\left(\frac{4(k_1^2+k_2^2)}{k_1^2k_2^2} \right) + O(\Lambda_z^{-1}) \right) .
\end{aligned}
\end{equation}
As before, this integral will include mixed terms of the form $f(x,y) \log(\Lambda_z)$ which cannot be regularized by a local counter term. 

In both these instances this diagnostic of spontaneous symmetry breaking agrees with the results in \cite{Seiberg:2020bhn,Seiberg:2020wsg,gorantla2021lowenergy}, namely that the symmetry is not spontaneously broken. Furthermore, these results confirm that the charged states have energies of order the lattice spacing and so correlation functions involving charged operators cannot be regularized to give a consistent IR result.

We would like to construct theories with a spontaneously broken continuous subsystem symmetry, as the IR physics of such theories would be described by the Goldstone modes of the broken symmetry. Finding such theories would be a first step to studying the phase transition between the broken and unbroken phases, a new type of critical phenomena that has yet to be considered in the literature. In the next two section we will analyze a family of statistical models inspired by the XY-plaquette model which do indeed have spontaneously broken subsystem symmetries.

\section{The classical XY-plaquette model} \label{sec:xy_plaquette}

To realize a theory with spontaneously broken subsystem symmetries, we can try to start from the lattice model related to the XY Plaquette model in $d$-dimensions \cite{paramekanti2002ring}. Consider a square lattice in $d$ dimensions with an angular variable $\phi_i$ on each lattice site. In the original XY model we consider the classical statistical mechanics model where the energy of each configuration of spins is given by
\begin{equation}
E_{\text{XY}}(\phi) = -J\sum_{\left<i,j\right>} \cos(\phi_i-\phi_j),
\end{equation}
where the sum runs over all sites $i,j$ which are nearest neighbors on the lattice. The partition function of this model is given by the path integral
\begin{equation}
Z(\beta) = \prod_i\int_0^{2\pi} d\phi_i ~~e^{-\beta E_{\text{XY}}(\phi)} .
\end{equation}

As the partition function is given by a Euclidean path integral, we can consider this model as a quantum mechanical model with a Euclidean action $S_E = E$, while the $d-1$ dimensional Hamiltonian will be related to this action by a Legendre transform. 

We can follow the exact same procedure to construct an XY-plaquette model by taking the interactions to be on plaquettes rather than between nearest neighbor sites. This will result in the energy 
\begin{equation} \label{eq:energy_plaquette}
E_{\text{XY-plaquette}}(\phi) = -J\sum_{S} \cos(\Delta_S \phi),
\end{equation}
where the sum runs over all square plaquettes $S$ containing the sites $\{s_1,s_2,s_3,s_4\}$ in cyclic order,  $\Delta_S\phi = \phi_{s_1} - \phi_{s_2} + \phi_{s_3} - \phi_{s_4}$, and $J$ is the interaction strength. The partition function of the statistical theory is described by the Euclidean path integral
\begin{equation}
Z(\beta) = \prod_i\int_0^{2\pi} d\phi_i ~~e^{-\beta E_{\text{XY-plaquette}}(\phi)} .
\end{equation}

This model is similar to the XY-plaquette model considered in \cite{paramekanti2002ring} and \cite{Seiberg:2020bhn}, as the plaquette interaction term is the same, though in our case $E_{\text{XY-plaquette}}(\phi)$ is the Euclidean action of the quantum model rather than the Hamiltonian, and it does not contain a kinetic term.

This XY-plaquette model has the subsystem symmetry of shifting all $\phi_i$'s on any $d-1$ dimensional hyper-plane in the lattice by a constant angle. Similar to the original XY model, we would expect this model to be in a broken phase at low energy if the dimension allows for such a phase to exist. The continuum limit of this broken phase should correspond to taking the continuum limit of the of the Euclidean action on the lattice. This was done carefully in \cite{Seiberg:2020bhn}, and the resulting model is a compact scalar field $\phi$ in $d$ dimensions with the Lagrangian
\begin{equation} \label{eq:lag_xy_plaquette}
L = \frac{1}{2g}\sum_{1 \leq i < j \leq d} \partial^{i} \partial^j \phi \partial_{i} \partial_j \phi .
\end{equation}

This Lagrangian has the global dipole symmetry of shifting $\phi$ by any arbitrary function of a single variable
\begin{equation} \label{eq:symmetry_xy_classical}
\phi \ra \phi + \sum_{i=1}^d f_i (x_i).
\end{equation}
As $\phi$ is $2\pi$ periodic, these functions are defined up to an integer multiple of $2\pi$. Thus neither $\phi$ nor $\partial_i \phi$ are good operators in the continuum limit, while $\partial_i \partial_j \phi$ is a good operator so long as $i \neq j$, as is $e^{i\phi}$, similar to the analysis of \cite{Seiberg:2020bhn}. Furthermore, we will only focus on field configurations that have a finite action in the continuum limit. This excludes any winding modes that have infinite action (due to a computation that is identical to the computation done in section 4.2 of \cite{Seiberg:2020bhn}).\footnote{In this model the momentum modes have finite action, as will be shown bellow, and so we can consistently exclude the winding modes from the continuum field theory. In \cite{Seiberg:2020bhn} both momentum and winding modes had infinite action in the continuum limit, and so one needs to keep both.}

 The spontaneous breaking of this dipole symmetry is what we are interested in analyzing. As this subsystem symmetry acts on $d-1$ dimensional hyper-planes, the minimal dimension we can hope to have a spontaneously broken symmetry is $d=4$, and indeed we will see that this is the case. However before we analyze the breaking of the dipole symmetry we present an analysis of the spectrum of the model. We note that a further study the correlation functions of uncharged operators is presented in appendix \ref{app:corr}.



\subsection{Spectrum} \label{sec:xy_plaq_spec}

To understand the spectrum of the model, we start from the equation of motion for $\phi$
\begin{equation}
    \sum_{1 \leq i < j \leq d} \partial_i^2 \partial_j^2 \phi = 0.
\end{equation}
Then we can expand $\phi$ in terms of Fourier modes, which leads to the dispersion relation
\begin{equation}
\sum_{1 \leq i < j \leq d} k_i^2k_j^2 = 0.
\end{equation}

If we consider the theory in Lorentzian signature, that is take one of the $d$ spatial coordinates and Wick rotate it, then the dispersion relation becomes
\begin{equation} \label{eq:dispersion}
    \omega^2 = \frac{\sum_{i\neq j}k_i^2k_i^2}{k^2}.
\end{equation}
We note that this dispersion relation is continuous around $k=0$, and indeed $\omega \rightarrow 0$ as $|k|\rightarrow 0 $ from any direction.  

Next we quantize the Lorenzian theory on a $d-1$-torus of equal length $ \ell$, so that the momenta are quantized $k_i = 2\pi \frac{n_i}{\ell}$. In terms of the momentum modes 
\begin{equation} \label{eq:phi_momnetummodes}
\phi_{\mathbf{n}} = \frac{1}{(2\pi)^{d-1}}\int d^{d-1}x ~\phi(x) e^{2\pi i \frac{\mathbf{n}\cdot \mathbf{x}}{\ell}} 
\end{equation}
the Lagrangian becomes
\begin{equation} \label{eq:lag_momentum_modes}
L = \frac{\ell^{d-3} (2\pi)^2}{2g}\sum_{\mathbf{n}\in \mathbb{Z}^3} \left[ |\mathbf{n}|^2 \partial_0 \phi_{-\mathbf{n}}\partial_0 \phi_{\mathbf{n}} 
- \frac{(2\pi)^2}{\ell^2}\sum_{i<j}n_i^2n_j^2\phi_{-\mathbf{n}}\phi_{\mathbf{n}} \right].
\end{equation}
Each of these modes for generic values on $\mathbf{n}$ behaves as a simple harmonic oscillator with ground state energy
\begin{equation}
E_{\mathbf{n}} = \frac{1}{2} \omega_{\mathbf{n}} = \frac{\pi}{\ell}\left(\frac{\sum_{i\neq j}n_i^2n_i^2}{\mathbf{n}^2}\right)^{1/2},
\end{equation}
where $\omega$ given by the dispersion relation \eqref{eq:dispersion}. The rest of the states are built via the generic Fock space construction on these ground states. If, however, all but one of the integers in $\mathbf{n}$ are zero then $\omega_{\mathbf{n}} = 0$ and these fields have no harmonic potential. We may need to quantize these modes more carefully, as they can have large momenta in the remaining direction. 

To quantize the Lagrangian for the modes where all but one of the $\mathbf{n}$'s are zero, we follow \cite{Seiberg:2020wsg} and write these modes in position space by writing the field as 
\begin{equation}
\phi(x_0,x_i,\ldots, x_{d-1}) = \phi_1(x_0,x_1) + \phi_2(x_0,x_2) + \ldots + \phi_{d-1}(x_0,x_{d-1}) + \ldots
\end{equation} 
Then the Lagrangian for these modes is
\begin{equation} 
    L = \frac{1}{2g} \sum_{i = 1}^{d-1} \left(\partial_0 \partial_i \phi_i\right)^2.
\end{equation}
We see that unlike in \cite{Seiberg:2020wsg} these modes are decoupled, and the ``gauge" symmetry that connects these modes is just a subset of the global symmetry of the Lagrangian $\phi \rightarrow \phi + f(t)$. Overall these modes behave just like the zero modes of a standard periodic free scalar field, and do not acquire a mass of the order of the lattice spacing, which is very different than the theories considered in \cite{Seiberg:2020bhn,Seiberg:2020wsg}.

This analysis shows that the even for the modes where all but one of the $\mathbf{n}$'s are zero the Lagrangian in \eqref{eq:lag_momentum_modes} is the correct Lagrangian, and the Hamiltonian for these modes in the $i$'th direction is
\begin{equation}
H_i = \frac{\ell^{d-3} (2\pi)^2}{2g} \sum_{n \neq 0} \frac{1}{n^2} \pi_{i,-n},\pi_{i,n} .
\end{equation}
The conjugate momenta of these modes, denoted by $\pi_{i,n}$, are the generators of the global symmetry $\phi \ra \phi + f_i(x_i)$. These conjugate momenta do indeed commute with the Hamiltonian, as expected for a global symmetry, though their spectrum is continuous. Choosing a ground state for the system is the same as choosing the eigenvalues of these modes, and so the $\pi_{i,n}$'s shift us from one ground state to another, in analogy to typical Goldstone modes. Thus this picture looks like a spontaneously broken symmetry, with the $\phi$ being an analogous Goldstone boson.

There is also the special mode $\mathbf{n} = 0$, for which the Lagrangian vanishes and so is not a physical mode. This is very different than in \cite{Seiberg:2020wsg} where this mode is the unique ground state of the system. The reason this mode disappears from the theory is the time translation symmetry $\phi \rightarrow \phi + f(t)$, which is generated by this zero mode. 

Finally we note is that the resulting Hamiltonian is non-local in space for all of the modes. Indeed for the continuum theory the conjugate variable to $\phi$ is 
\begin{equation}
\pi = \frac{\partial L}{\partial(\partial_0\phi)} = \frac{1}{g}\partial_0 \nabla^2\phi,
\end{equation}
so the Hamiltonian of the system is 
\begin{equation}
    H = \int d^{d-1} x~\left[ \frac{1}{2g}\pi \nabla^{-2} \pi  + \frac{1}{2g}\sum_{1 \leq i < j \leq d-1} \left(\partial_i \partial_j \phi\right)^2 \right],
\end{equation}
which has a non-local kinetic term. This Hamiltonian still has a fine expansion in terms of Fourier modes, as seen above, and so can be quantized in the standard way. 

It is interesting that a local Lagrangian gave rise to a non-local Hamiltonian, which raises the question which is more fundamental. From an effective field theory prospective, one writes down the most general local Lagrangian that obeys the desired symmetries. Then one keeps only the renormalizable terms, which are found by computing the scaling dimensions. Based on this philosophy a local Lagrangian is more fundamental to understanding the IR physics, as it contains the relevant information about the symmetries and scaling dimensions. The model we presented fits nicely into this picture, as it is the simplest Lagrangian that obeys the desired subsystem symmetry, and indeed it seems that the resulting quantum field theory is local and well defined.

There is however the alternate bottom up approach where one starts from a local UV Hamiltonian, say on a lattice, and then tries to infer the physics at large distances. In this case one assumes that the Hamiltonian is local, and so it does not seem that the classical XY model can describe the IR physics that emerge from such a construction. This raises many interesting questions relating to the existence of phases of matter that fit into an effective field theory picture, but that cannot be described by a local lattice Hamiltonian. 

There is also the possibility that certain phases of matter may arise from a local lattice Hamiltonian that has a non-local Lagrangian. For example one can introduce a nearest neighbor interaction of the conjugate momenta variables to a lattice Hamiltonian, which would keep the Hamiltonian local but result in a non-local Lagrangian. Such systems have been considered in the condensed matter literature, for example \cite{Wu_2021} studied similar Hamiltonians with subsystem symmetries, though it is not clear to us how such models fit into the canonical picture of Wilsonian quantum field theory.

\subsection{Spontaneous subsystem symmetry breaking}





To identify if the subsystem symmetry  \eqref{eq:symmetry_xy_classical} is spontaneously broken, we will look at long range correlation functions of the charged local operator $e^{i \phi}$.\footnote{We explore the correlation functions of uncharged operators and their consequences in appendix \ref{app:corr}.} The 1, 2, and 3-point functions of these operators must vanish by the global dipole symmetry. Hence the simplest nonzero correlation function of such operators is the 4-point rectangle function, similar to what was considered in \eqref{eq:corr_xy_org}. This correlator is
\begin{equation} \label{eq:corr_xy_classical}
\left<e^{i [\phi(0,0) -\phi(x_1,0)-\phi(0,x_2) + \phi(x_1,x_2)]} \right>,
\end{equation}
where we have chosen the rectangle to lie in the $(x_1,x_2)$ plane, and suppressed the dependence on the remaining $d-2$ coordinates.

Taking the Gaussian functional integral we can write this expectation value as
\begin{equation}
\left<e^{i [\phi(0,0) -\phi(x_1,0)-\phi(0,x_2) + \phi(x_1,x_2)]} \right>
 = e^{-2g I_d(x_1,x_2)},
\end{equation}
where
\begin{equation}
I_d(x,y) = \int \frac{d^d k}{(2\pi)^d} \frac{(1-\cos(k_1x))(1-\cos(k_2y))}{\sum_{i\neq j} k_i^2 k_j^2}.
\end{equation}
Notice that this integral is well defined in the IR, as around $k_1,k_2 \ll 1$ this integral behave as $(k_1^2 k_2^2)/(\sum_{i\neq j}k_i^2 k_j^2)$, which is finite around zero for a fixed angle. 

For $d=2$ we can take this integral analytically, and we find
\begin{equation}
I_2(x,y) = \frac{|xy|}{4}. 
\end{equation}
This grows as we take $x$ and $y$ to be large, hence the full correlator \eqref{eq:corr_xy_classical} vanishes, and thus the subsystem symmetry is unbroken.

For $d=3$ this integral evaluates to
\begin{equation}
\begin{aligned}
I_3(x,y) &= \frac{1}{4\pi} \left[|y|\log\left(\frac{|x|+\sqrt{x^2+y^2}}{|y|} \right) + |x|\log\left(\frac{|y|+\sqrt{x^2+y^2}}{|x|} \right) \right] \\
& = \frac{1}{4\pi} |xy|^{1/2} \left\{\alpha^{-1/2}\log\left(\alpha +\sqrt{1+\alpha^2} \right) + \alpha^{1/2} \log\left(\frac{1 +\sqrt{1+\alpha^{2}}}{\alpha} \right) \right\},
\end{aligned}
\end{equation}
where $\alpha \equiv \frac{|x|}{|y|}$. Again the integral grows as $x$ or $y$ (or both) are taken to be large, implying that the subsystem symmetry is still unbroken. This is expected as the minimal dimension allowed for spontaneous symmetry breaking in this model is $d=4$.

For $d=4$ we can write the rectangular function as 
\begin{equation}
\begin{aligned}
I_4(x,y) &= \frac{1}{8\pi^3} \int dk_1 ~dk_2 ~ \frac{(1-\cos(k_1x))(1 - \cos(k_2y) )}{k_1^2+k_2^2} K\left( 1 - \frac{k_1^2k_2^2}{(k_1^2+k_2^2)^2}\right) ,
\end{aligned}
\end{equation}
where $K(m)$ is the complete elliptic integral of the first kind. Notice that $I_4(x,y)$ is only a function of the ratio $ x/y$, and so we expect it to be a constant for large rectangles with the ratio $x/y$ fixed. However this integral is logarithmically divergent in the UV, and this UV divergence neads to be regularized. Moving to polar coordinates and introducing a UV cuttoff, we can take the radial integral to find
\begin{equation}
\begin{aligned}
I_4(x,y) &= \frac{1}{8\pi^3} \int_0^\Lambda dr \int_0^{2\pi} d\theta ~ \frac{(1-\cos(x r \cos \theta))(1 - \cos(yr \sin \theta) )}{r} K\left( 1 - \cos^2 \theta \sin^2 \theta\right) \\
& = \frac{1}{8\pi^3}  \int_0^{2\pi} d\theta ~ K\left( 1 - \cos^2 \theta \sin^2 \theta\right) \left[\log \Lambda + \gamma +\frac{1}{2}\log \left(\frac{x^2 y^2 \cos^2 \theta \sin^2\theta}{\left|x^2\cos^2\theta - y^2 \sin^2\theta \right|} \right) \right] + O(\Lambda^{-1}).
\end{aligned}
\end{equation}
Notice that the angular integral multiplying the $\log\Lambda$ term is finite, hence we can renormalize the operator by a local counterterm to cancel the logarithmic UV divergence, leaving the remaining integral finite and cutoff independent. Canceling the UV cutoff does however make the integral logarithmically diverge as the area of the rectangles is taken to be large while fixing the ratio of the two sides. However this term approaches a constant if only one side of the rectangle is taken to be large while the other is kept fixed (notice that this is not the case for $I_2$ and $I_3$ which both diverge even in this limit, though slower than the divergence for fixed ratio). This implies that the subsystem symmetry is indeed spontaneously broken for $d=4$. 

\begin{figure}
    \centering
    \includegraphics[width = 14 cm]{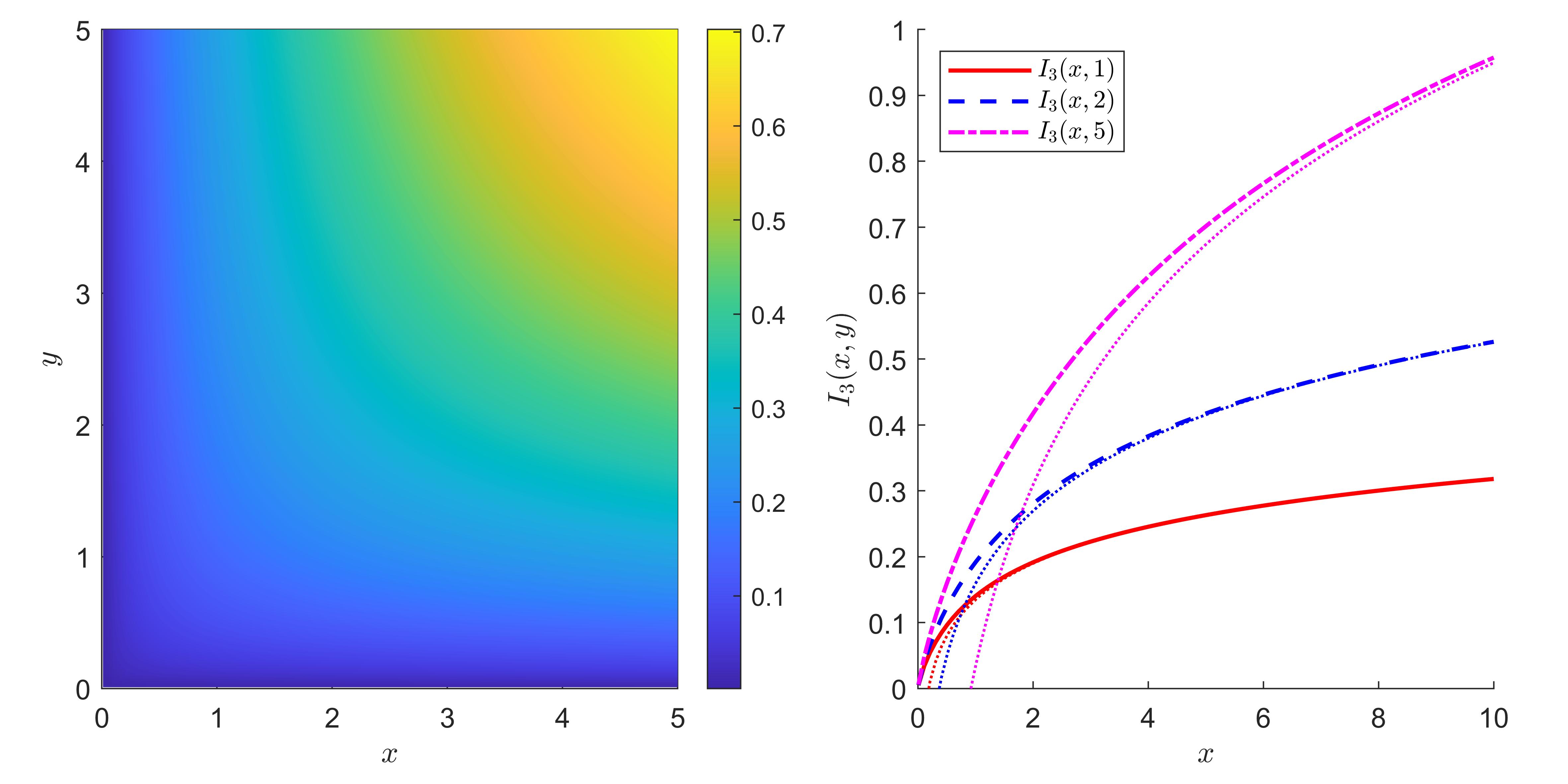}
    \caption{A surface plot of $I_3(x,y)$, as well as a plot of $I_3(x,y)$ for fixed values of $y$. We also compare $I_3(x,y)$ at fixed $y$ to the asymptotic behavior at large $x$ denoted by the dotted lines. The asymptotic behavior is $I_3(x,y) \approx \frac{y}{4\pi} \left(\log(2x/y) + 1 + O(1/x^2) \right)$. }
    \label{fig:I_3}
\end{figure}

\begin{figure}
    \centering
    \includegraphics[width = 14 cm]{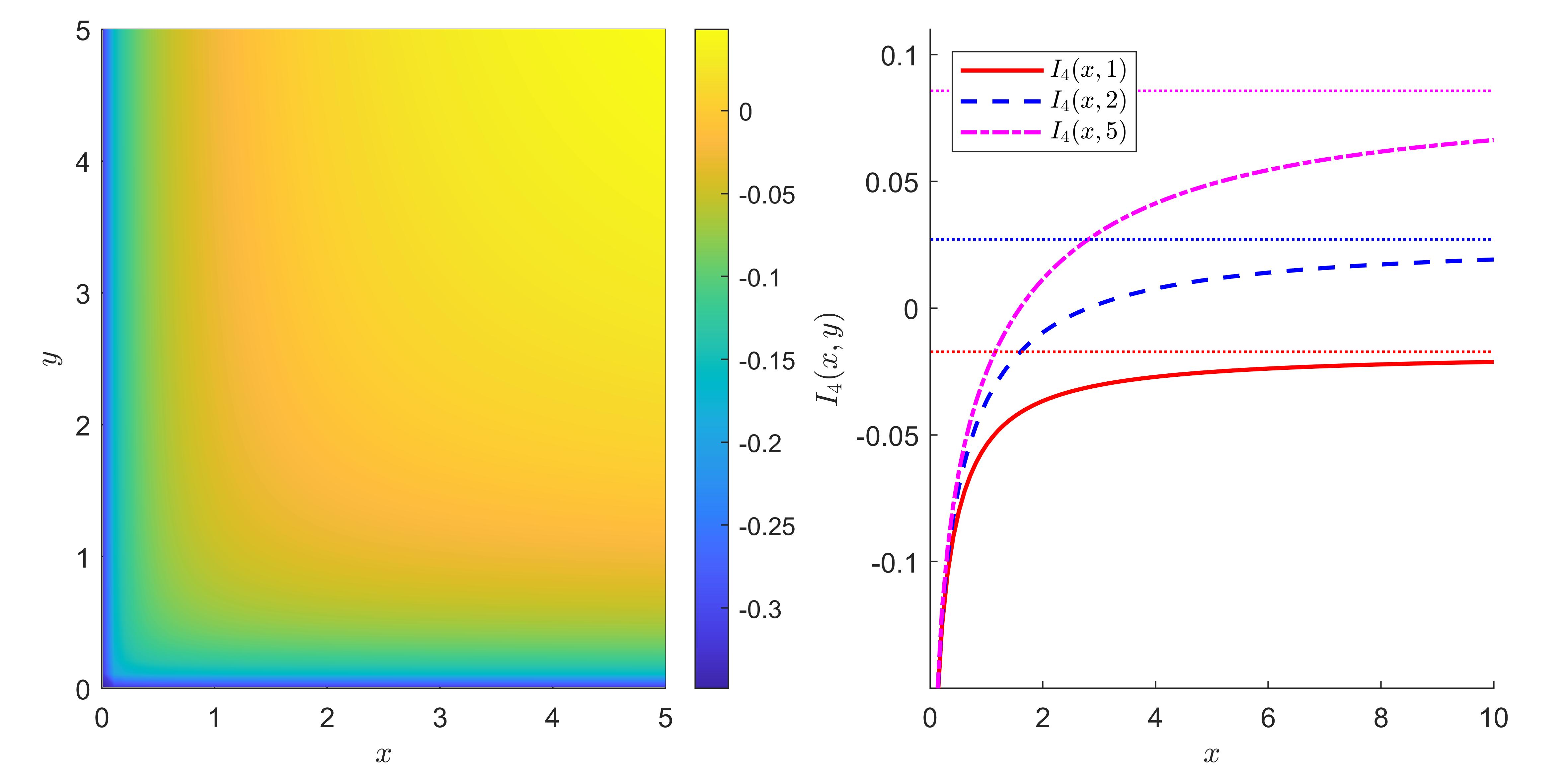}
    \caption{A surface plot of $I_4(x,y)$, as well as a plot of $I_4(x,y)$ for fixed values of $y$. We also compare $I_4(x,y)$ at fixed $y$ to the asymptotic value at infinite $x$ denoted by the dotted lines.}
    \label{fig:I_4}
\end{figure}

We expect a similar behavior in dimensions $d>4$, where an operator renormalization can take care of the UV divergence, while no IR divergent will exist at all. This is in alignment with the previous discussions of spontaneously breaking subsystem symmetries based on \cite{Batista:2004sc}. For completeness we present plots of $I_3(x,y)$ and $I_4(x,y)$ in figures \ref{fig:I_3} and \ref{fig:I_4}.

\section{Other classical XY models with subsystem symmetries}

The classical XY-plaquette model can be generalized to a family of models in $d$ dimensions that has a subsystem symmetry which acts on $d-m+1$ dimensional hyperplanes. As before we start with a square lattice in $d$ dimensions with an angular variable $\phi_i$ on each lattice site, though now the energy of each configuration is given by 
\begin{equation} \label{eq:energy_hypercube}
E(\phi) = -J\sum_{C_m} \cos(\Delta_{C_m} \phi).
\end{equation} 
This the sum runs over all hyper-cubes $C_m$ containing the sites $\{c_{s+\sum_{j=1}^m i_j \mathbf{e}_j}\}$, where $s$ is any lattice site, $i_j\in\{0,1\}$, and $\mathbf{e}_m$ are $m$ orthogonal unit vectors on the lattice. The classical XY model is the case $m=1$, while the classical XY-plaquette model is the case $m=2$. 

The next class of models is the case $m=3$, which is a classical version of the XY-cube model considered in \cite{You_2020,Gorantla:2020xap}. We present a brief analysis of this model below, before considering the general $m$ case.

\subsection{The classical XY-cube model}

The classical XY-cube model has the subsystem symmetry of shifting all $\phi_i$'s on any $d-2$ dimensional hyper-plane in the lattice by a constant angle. Similar to the previous model, we would expect this model to be in a broken phase at low energy if the dimension allows for such a phase to exist, which for this model is $d>4$. The continuum limit of this broken phase should correspond to taking the continuum limit of the of the Euclidean action on the lattice. This was done carefully in \cite{Gorantla:2020xap}, and the resulting model is a compact scalar field $\phi$ in $d$ dimensions with the Lagrangian
\begin{equation}
L = \frac{1}{2g}\sum_{1 \leq i < j < k \leq d} \partial^{i} \partial^j \partial^k \phi \partial_{i} \partial_j \partial_k \phi .
\end{equation}

This Lagrangian has the global dipole symmetry of shifting $\phi$ by any arbitrary function of any two variables
\begin{equation}
\phi \ra \phi + \sum_{1 \leq i < j \leq d} f_{i,j} (x_i,x_j).
\end{equation}

As $\phi$ is $2\pi$ periodic, these functions are defined up to an integer multiple of $2\pi$. Thus neither $\phi$ nor $\partial_i \phi$ or $\partial_i \partial_j \phi$ are good operators in the continuum limit, while $\partial_i \partial_j \partial_k \phi$ is a good operator so long as $i \neq j \neq k$, as is $e^{i\phi}$, similar to the analysis of \cite{Gorantla:2020xap}. As before, we will only focus on field configurations that have a finite action in the continuum limit, excluding any winding modes that have infinite action.

To find the spectrum of the theory, we can expand $\phi$ in terms of Fourier modes, which leads to the dispersion relation
\begin{equation}
\sum_{1 \leq i < j<l \leq d} k_i^2k_j^2 k_l^2 = 0.
\end{equation}
If we consider the theory in Lorentzian signature, that is take one of the $d$ spatial coordinates and Wick rotate it, then the dispersion relation becomes
\begin{equation} \label{eq:dispersion_cube}
    \omega^2 = \frac{\sum_{i< j<l}k_i^2k_k^2 k_l^2}{\sum_{i<j}k_i^2 k_j^2}.
\end{equation}
As before, this dispersion relation is continuous around $k=0$.

We can quantize the Lorenzian theory on a $d-1$-torus of equal length $ \ell$, so that the momenta are quantized $k_i = 2\pi \frac{n_i}{\ell}$. In terms of the momentum modes given in \eqref{eq:phi_momnetummodes}, the Lagrangian becomes
\begin{equation} \label{eq:lag_momentum_modes_cube}
L = \frac{\ell^{d-5} (2\pi)^4}{2g}\sum_{\mathbf{n}\in \mathbb{Z}^3} \left[ \left( \sum_{i<j}n_i^2n_j^2\right) \partial_0 \phi_{-\mathbf{n}}\partial_0 \phi_{\mathbf{n}} 
- \frac{(2\pi)^2}{\ell^2}\sum_{i<j<k}n_i^2n_j^2n_k^2\phi_{-\mathbf{n}}\phi_{\mathbf{n}} \right].
\end{equation}
Each of these modes for generic values on $\mathbf{n}$ behaves as a simple harmonic oscillator with ground state energy
\begin{equation}
E_{\mathbf{n}} = \frac{1}{2} \omega_{\mathbf{n}} = \frac{\pi}{\ell}\left(\frac{\sum_{i< j<k} n_i^2 n_j^2 n_k^2}{\sum_{i<j}n_i^2 n_j^2}\right)^{1/2},
\end{equation}
with $\omega$ given by the dispersion relation \eqref{eq:dispersion_cube}. The rest of the states are built via the generic Fock space construction on these ground states. If, however, all but two of the integers in $\mathbf{n}$ are zero then $\omega = 0$ and these modes have no harmonic potential. As before, these modes behave just like the zero modes of a standard periodic free scalar field, and do not acquire a mass of the order of the lattice spacing, which is very different than the theories considered in \cite{Seiberg:2020bhn,Seiberg:2020wsg,Gorantla:2020xap}. 

This analysis shows that, even for the modes where all but two of the $\mathbf{n}$'s are zero, the Lagrangian in \eqref{eq:lag_momentum_modes_cube} is correct. The momenta conjugate to these modes, denoted by $\pi_{i,j,n,m}$, are the generators of the global symmetries $\phi \ra \phi + f_{ij}(x_i,x_j)$. These conjugate momenta do indeed commute with the Hamiltonian, as expected for a global symmetry. Choosing a ground state for the system is the same as choosing the eigenvalues of these operators, and so the $\pi_{i,j,n,m}$'s shift us from one ground state to another, just as the zero modes for the usual Goldstone boson. The modes corresponding to where all but one of the $n$'s are zero now have vanishing action, and correspond to the symmetries involving the time direction $\phi \ra \phi + f_{0j}(x_0,x_j)$.  Finally we note that the resulting Hamiltonian is non-local in space, and the discussion regarding this non-locality from section \ref{sec:xy_plaq_spec} applies to this model as well.


As in the XY-Plaquette model, to check for which dimensions the symmetry is spontaneously broken we need to compute correlation functions of charged operators. In this model the simplest invariant correlation function is the cubic correlator of eight $e^{i\phi}$ given by
\begin{equation}
    \left<\exp\left[i\sum_{i,j,k \in \{0,1\}} (-1)^{i+j+k}\phi(ix,jy,kz) \right]\right> .
\end{equation}

As this theory is Gaussian, we can take the path integral directly resulting in
\begin{equation}
    \left<\exp\left[i\sum_{i,j,k \in \{0,1\}} (-1)^{i+j+k}\phi(ix,jy,kz) \right]\right>  = e^{- 4g J_d(x,y,z)}
\end{equation}
where
\begin{equation}
    J_d(x,y,z) = \int \frac{d^d k}{(2\pi)^d} \frac{(1-\cos(k_1x))(1-\cos(k_2y))(1-\cos(k_3z))}{\sum_{i< j<l} k_i^2 k_j^2 k_l^2}.
\end{equation}

For $d=3$, the lowest dimension for which the model makes sense, we have that
\begin{equation}
    J_3(x,y,z) = \frac{1}{8}|xyz|
\end{equation}
As before, this implies that the symmetry is not broken as the long range correlation functions grows.

For $d=4$ this integral is
\begin{equation}
    J_4(x,y,z) =  \int \frac{d^3 k}{16 \pi^3} \frac{(1-\cos(k_1x))(1-\cos(k_2y))(1-\cos(k_3z))}{|k_1k_2k_3|\sqrt{k_1^2k_2^2+k_2^2k_3^2+k_1^2k_3^2}} .
\end{equation}

This integral is completely convergent for any finite values of $x,y,z$, though a closed form expression for the integrand is beyond our reach. We can however take various large distance limits of this function. Just via scaling arguments it is clear that if we take $x,y$ and $z$ to be large while keeping their ratios fixed then
\begin{equation}
    J_4(x \ra \infty, y \ra \infty, z \ra \infty) = |xyz|^{2/3} f(x/y,x/z),
\end{equation}
where
\begin{equation}
    f(a,b) = \int \frac{d^3 k}{16 \pi^3} \frac{(1-\cos(k_1))(1-\cos(k_2))(1-\cos(k_3))}{|k_1k_2k_3|\sqrt{a^{2/3}b^{-4/3} k_1^2k_2^2+ a^{2/3}b^{2/3} k_2^2k_3^2+ a^{-4/3} b^{2/3} k_1^2k_3^2}}.
\end{equation}

If we take only $x$ and $y$ to be large while keeping $z$ and the ratio $x/y$ fixed then
\begin{equation} \label{eq:J_4_xyinf}
\begin{aligned}
    J_4(x \ra \infty, y \ra \infty, z) &= \frac{1}{4}|xy|^{1/2} |z| \int \frac{d^2 k}{(2\pi)^2} \frac{(1-\cos k_1)(1-\cos k_2)}{|k_1 k_2| \sqrt{(y/x)k_1^2 + (x/y)k_2^2}} \\
    &= \frac{|z| |xy|^{1/2}}{8\pi}  \left\{\alpha^{-1/2}\log\left(\alpha +\sqrt{1+\alpha^2} \right) + \alpha^{1/2} \log\left(\frac{1 +\sqrt{1+\alpha^{2}}}{\alpha} \right) \right\},
\end{aligned}
\end{equation}
where $\alpha \equiv |x/y|$.

Finally, taking $x\ra \infty$ while keeping $y$ and $z$ fixed we see that the integral scales as 
\begin{equation}
    J_4(x \ra \infty, y,z) = \frac{1}{8\pi} |y z| \big(\log|x| + O(1) \big),
\end{equation}
so even in this limit the cube correlation function diverges. As this symmetry acts on $d-2$ dimensional hyper-planes, it cannot be spontaneously broken in dimensions $d\leq 4$, so this analysis does in fact make sense. 

For this symmetry to be spontaneously broken we need to be in at least $d=5$, where the integral evaluates to 
\begin{equation}
\begin{aligned}
    J_5(x,y,z) &= \int \frac{d^4k}{2 (2\pi)^4} \frac{(1-\cos(k_1 x))(1-\cos(k_2y))(1-\cos(k_3z))}{ \sqrt{\left( \sum_{i<j<l} k_i^2k_j^2k_l^2 \right) \left(\sum_{i<j} k_i^2 k_j^2 \right) }} \\
    &= \int \frac{d^3k}{(2\pi)^4} \frac{(1-\cos(k_1x))(1-\cos(k_2y))(1-\cos(k_3z))}{|k_1 k_2 k_3| \sqrt{k_1^2+k_2^2+k_3^2} } K\left(1 - \frac{\left(k_1^2k_2^2+ k_1^2k_3^2 + k_2^2 k_3^2\right)^2}{k_1^2k_2^2k_3^2 \left( k_1^2+k_2^2+k_3^2 \right) }\right) .
\end{aligned} 
\end{equation}
This integral is still UV finite, and converges for any values of $x,y$ and $z$, though it has no known closed form expression. We can however do the same limit  analysis to how it behaves when we take the cube to be large. If we take $x \ra \infty$ while keeping $y$ and $z$ fixed then the integral approximates as \footnote{We can write $J_5 = \int d k_1 (1-\cos(k_1))f(k_1)$, and as $f(k_1)$ is an integrable function its Fourier transform decays at infinity. Thus the integral $\int dk_1 \cos(k_1 x) f(k_1)$ vanishes as $x\ra \infty$ and we are left with the remaining finite integral.}
\begin{equation} \label{eq:J_5_xinf}
\begin{aligned}
    J_5(x \ra \infty,y,z) = 2\pi \int d^3k \frac{(1-\cos(k_2y))(1-\cos(k_3z))}{|k_1 k_2 k_3| \sqrt{k_1^2+k_2^2+k_3^2} } K\left(1 - \frac{\left(k_1^2k_2^2+ k_1^2k_3^2 + k_2^2 k_3^2\right)^2}{k_1^2k_2^2k_3^2 \left( k_1^2+k_2^2+k_3^2 \right) }\right),
\end{aligned}
\end{equation}
which is a constant value independent of $x$. Hence this symmetry is indeed broken in $d=5$, as expected.

\subsection{Classical XY-hypercube models}

We now turn our attention back to the general $m$ case, the classical XY-hypercube model of degree $m$, with the energy given by \eqref{eq:energy_hypercube}. This model has a subsystem symmetry which acts on $d-m+1$ dimensional hyperplanes. The continuum field theory of these models consists of a periodic scalar field $\phi$ with the Lagrangian
\begin{equation}
    L = \frac{1}{2g}\sum_{i_1 < i_2< \ldots <i_m = 1}^d \partial^{i_1}\partial^{i_2}\ldots  \partial^{i_m} \phi ~ \partial^{i_1}\partial^{i_2}\dots  \partial^{i_m} \phi .
\end{equation}

This model has the subsystem shift symmetry given by
\begin{equation} \label{eq:symmetry_general_m}
    \phi \ra \phi + f_{i_1,i_2,\ldots, i_{m-1}}(x_{i_1},x_{i_2},\ldots,x_{i_{m-1}})
\end{equation}
which acts on $d-m+1$ dimensional hyperplanes. 

The spectrum of the model with arbitrary $m$ has many of the same features as the spectrum of the plaquette and cubic models. The dispersion relation in Lorentzian signature will be given by
\begin{equation} \label{eq:dispertion_general_m}
    \omega^2 = \frac{\sum_{i_1 < \ldots <i_{m}} k_{i_1}^2 \ldots k_{i_{m}}^2}{\sum_{i_1 < \ldots <i_{m-1}} k_{i_1}^2 \ldots k_{i_{m-1}}^2} ,
\end{equation}

The quntization of the theory on a torus follows the same procedure as was done for the plaquette and cubic models. Generic modes will behave like Harmonic oscillators with ground state energy $E = \frac{1}{2}\omega$, where $\omega$ is given by the dispersion relation \eqref{eq:dispertion_general_m}. Modes where all but $m-1$ of the $k$'s have  zero harmonic potential but a non-zero kinetic term, and so are the zero modes of the theory. These modes generate the subset of shift symmetries in \eqref{eq:symmetry_general_m} that do not involve the time direction. Similarly, the modes where all but $m-2$ of the $k$'s are zero have a vanishing Lagrangian and generate the symmetries that do involve the time direction. 

The dimensions for which the subsystem symmetry \eqref{eq:symmetry_general_m} can be spontaneously broken is $d>1+m$ (by the generalized Merman-Wagner-Coleman theorem \cite{Batista:2004sc}). We have shown that the symmetry is indeed spontaneously broken when $d=2+m$ for $m\leq 3$\footnote{We considered $m=2,3$ in this paper, while the case $m=1$ is just a free compact scalar field which is known to exhibit spontaneous symmetry breaking in $d=3$.}, and expect this trend to continue for arbitrary $m$.

To diagnose whether the subsystem symmetry \eqref{eq:symmetry_general_m} is spontaneously broken, one would need to compute a correlation function of $e^{\pm i\phi}$ operators placed on the corners of a $m$-dimensional hypercube, as this is the simplest correlation function of charged operators that is uncharged under the subsystem symmetry. Calling this correlation function $F_m(x_1,x_2,\ldots,x_m)$, one can preform the Gaussian path integral to find that
\begin{equation}
    F_m(x_1,x_2,\ldots,x_m) = e^{-2^{m-1}g\; I_{m,d}},
\end{equation}
where 
\begin{equation} \label{eq:I_m_d}
    I_{m,d}(x_1,x_2,\ldots,x_m) = \int \frac{d^dk}{(2\pi)^d} \frac{\prod_{i=1}^m(1 - \cos(k_i x_i) )}{\sum_{i_1 < \ldots <i_{m}} k_{i_1}^2 \ldots k_{i_{m}}^2} .
\end{equation}
The behavior of $I_{m,d}(x_1,x_2,\ldots,x_m)$ at large separations determines the spontaneous breaking of the subsystem symmetry; if $I_{m,d}$ grows at large separation then the symmetry is unbroken while if it approaches a constant then the symmetry is spontaneously broken. We have shown that $I_{2,4}$ and $I_{3,5}$ approach a constant when one of the coordinates is taken to be large, and expect $I_{m,m+2}$ to behave the same way for arbitrary $m$.

We can actually compute this limit directly from \eqref{eq:I_m_d}. Notice that $I_{m,m}$ factorizes, and thus we can directly compute
\begin{equation} 
    I_{m,m}(x_1,x_2,\ldots,x_m) = 2^{-m}|x_1 x_2 \ldots x_m| .
\end{equation}
Hence the symmetry is not spontaneously broken for $d=m$. 

For $d=m+1$ we can take the limit $x_1 \ra \infty$ while keeping the rest of the variables fixed,\footnote{The simplest way to do this calculation is to first take both $x_1$ and $x_2$ to be large, when the limit reduces to an integral similar to \eqref{eq:J_4_xyinf}. Then one can take the limit $x_1\ra \infty$ while keeping $x_2$ fixed and obtain \eqref{eq:I_m_m1}.} and we find that
\begin{equation} \label{eq:I_m_m1}
    I_{m,m+1}(x_1 \ra \infty ,x_2,\ldots,x_m) = 2^{-m} \pi^{-1}|x_2 \ldots x_m| \log(x_1),
\end{equation}
again implying the symmetry is not spontaneously broken.

For $d=m+2$ the integral $I_{m,m+2}$ is UV finite assuming $m>2$, so we can take the same limit $x_1 \ra \infty$ while keeping the rest of the variables fixed, resulting in the finite value \footnote{This limit is obtained in the same way as \eqref{eq:J_5_xinf}.}
\begin{equation}
\begin{aligned}
    I_{m,m+2}(x_1 \ra \infty, x_2,\ldots,x_m) = \int & \frac{d^m k}{(2\pi)^{m+1}}  \frac{\prod_{i=2}^m(1-\cos(k_ix_i))}{ \prod_{i=1}^m |k_i|\left( \sum_{i_1<\ldots<i_{m-2}}k_{i_1}^2 \ldots k_{i_{m-2}}^2 \right)^{1/2} } \times\\
    & \qquad \times 
    K\left(1 - \frac{\left( \sum_{i_1<\ldots<i_{m-1}}k_{i_1}^2 \ldots k_{i_{m-1}}^2 \right)^2}{\prod_{i=1}^m k_i^2\left(  \sum_{i_1<\ldots<i_{m-2}}k_{i_1}^2 \ldots k_{i_{m-2}}^2 \right) }\right) . 
\end{aligned}
\end{equation}
As in this limit $I_{m,m+2}$ approaches a constant (i.e. is independent of $x_1$) the symmetry is indeed spontaneously broken. Hence the subsystem symmetry \eqref{eq:symmetry_general_m} is spontaneously broken when $d=m+2$ for arbitrary $m$, but not when $d \leq m+1$. This is in agreement with the expectations based on the generalized Merman-Wagner-Coleman theorem \cite{Batista:2004sc}.

\section{Discussion and future directions}

Our main goal in this paper was to discuss spontaneous breaking of subsystem symmetries, and to construct well behaved continuum quantum field theories that model the spontaneously broken phase. Though we successfully built field theories that exhibit a spontaneous broken subsystem symmetry, there are still many open questions relating to these models, and to quantum field theories with subsystem symmetries in general. In this section we discuss in further detail two general areas of study that we find most interesting. The first is to study the unbroken phase and the phase transition in the classical XY-plaquette model we constructed in section \ref{sec:xy_plaquette}. The second is to analyze generalizations and interacting theories built upon the classical XY-plaquette model.

There are however many other interesting aspects to keep investigating with regards to the models we constructed. One natural question to ask is whether the subsystem symmetries are still spontaneously broken at finite temperature, assuming such breaking is allowed by the number of dimensions. As our models have no free parameters, the symmetry is either broken or unbroken at all finite temperatures. To understand which is the case, we need to compute the same 4-point correlation functions of exponential operators we considered when taking one of the spatial dimensions to be compact. Though one can evaluate the sum over the Matsubara frequencies in the compact direction, the resulting integrals over the momenta in the remaining directions are unwieldy, and must be UV regularized for the classical XY-plaquette model. Hence we decided to leave this problem for a future endeavour.

Another interesting observation is that the Lagrangian \eqref{eq:lag_xy_plaquette} in $d=2$ is similar to the Lagrangian of a zero velocity chiral boson, $\mathcal{L} = \alpha \partial_0 \phi \partial_1 \phi$. The chiral boson also possesses the same subsystem symmetry \eqref{eq:symmetry_xy_classical}, which is not spontaneously broken in $d=2$. One can try to generalize the chiral Lagrangian to higher dimensions, namely consider the Lagrangian $\mathcal{L} = \alpha \sum_{i\neq j}\partial_i \phi \partial_j \phi$ which also possesses the same subsystem symmetry. However this Lagrangian breaks the reflection symmetry of the lattice, and also has a Hamiltonian that is unbounded from below. One can overcome some of the difficulties in quantizing the $d=2$ theory using the conformal symmetry, and perhaps in higher dimensions there is a different conformal symmetry which emerges, similar to the ones studied in \cite{Karch_2021}, which may be worth studying.

\subsection{The unbroken phase and phase transition}

Up to now we have focused on the symmetry broken phases of the classical XY models we considered. However, just like the standard XY model, the lattice models have a phase where the symmetry is unbroken, and a phase transition between the two phases. To model the phase transition and the unbroken phase, we would like to construct a Lagrangian for a complex scalar field $\Phi$ that has the desired $U(1)$ subsystem symmetry. Then tuning the potential of $\Phi$ in this Lagrangian would (at least classically) give $\Phi$ a vacuum expectation value, and thus take us between the spontaneously broken and unbroken phases.

This Lagrangian can be built in a similar manner to the Lagrangian's with global dipole symmetries constructed in \cite{Pretko_2018,Gromov2019}. Focusing on the classical XY-plaquette model, the lowest order terms in this effective Lagrangian are
\begin{equation} \label{eq:lag_XY_complex}
    L = \sum_{1\leq i<j \leq d}\left|\Phi\partial_i\partial_j \Phi - \partial_i\Phi \partial_j\Phi \right|^2 + V(|\Phi|^2) .
\end{equation}
In the symmetry broken phase we would choose the potential in \eqref{eq:lag_XY_complex} to be a Mexican hat. We can then expand the Lagrangian around one of the minima of the potential, and to leading order in the fields we would arrive at the continuum Lagrangian for the XY-plaquette model given in equation \eqref{eq:lag_xy_plaquette}. This is a simple check that the Lagrangian we wrote down has the desired symmetry broken phase that we analysed in section \ref{sec:xy_plaquette}.

To access the unbroken phase one would need to choose a generic potential with a minimum at $|\Phi|^2 = 0$. Around this minimum the Lagrangian to lowest orders reads
\begin{equation} \label{eq:lag_XY_complex_broken}
    L = \sum_{1\leq i<j \leq d}\left|\Phi\partial_i\partial_j \Phi - \partial_i\Phi \partial_j\Phi \right|^2 + m^2|\Phi|^2 + \lambda |\Phi|^4 .
\end{equation}
Unfortunately this effective Lagrangian is non-Gaussian even at lowest order due to the kinetic term containing 4 $\Phi$ fields. Hence an analytic or even perturbative description of the dynamics in the unbroken phase is beyond our reach (as was also the case in \cite{Pretko_2018}.) This phase can still be studied numerically, either on the lattice using \eqref{eq:energy_plaquette} or in the continuum field theory \eqref{eq:lag_XY_complex_broken}. This phase may also be accessible analytically using non-perturbative techniques, like perhaps taking some sort of large $N$ limit. 

It would also be of great interest to study the phase transition between broken and unbroken phases. However, as the effective Lagrangian \eqref{eq:lag_XY_complex_broken} is non-Gaussian and strongly coupled in the vicinity of the phase transition, even a mean field approximation of the transition is not possible. It is not even clear if the phase transition is continuous or first order, and the literature on phase transitions in similar models is uncertain. 

For example the plaquette Ising model \cite{suzuki1972solution}, which is the $\mathbb{Z}^2$ version of the classical XY-plaquette model \eqref{eq:energy_plaquette}, has been shown to undergo a weakly first order phase transition \cite{mueller2014nonstandard,johnston2017plaquette}. On the other hand a recent study of the phase transition in models related to the $\mathbb{Z}^N$ versions of the X-cube model indicated that the transition is continuous for $N>4$ \cite{lake2021subdimensional}. As the plaquette Ising model is dual to the $\mathbb{Z}^2$ X-cube model \cite{Vijay_2016}, these two results are in agreement. One may then think of the XY-plaquette model as a $N\rightarrow \infty$ limit of $\mathbb{Z}^N$ models, in which case the phase transition is likely continuous, though more numerical or analytical evidence is needed to confirm this conjecture.

\subsection{Generalizations and interacting theories}

Aside from modelling the unbroken phase and the phase transition, it would also be interesting to construct interacting continuum field theories based on the classical XY-plaquette model. We present two general systematic constructions of such theories: the first method is to gauge the $U(1)$ subsystem symmetry to a tensor gauge theory in the spirit of \cite{Pretko_2018}, while the second idea is to promote a global $U(1)$ symmetry to a subdimentional symmetry.

The most straightforward generalization of the classical XY-plaquette model is to gauge the symmetry \eqref{eq:symmetry_xy_classical}. This can be done by starting with the generalized Lagrangian \ref{eq:lag_XY_complex}, and then replacing the derivative term $\Phi \partial_i \partial_j \Phi - \partial_i\Phi \partial_j \Phi$ with a ``covariant" derivative term similar to \cite{Pretko_2018}, 
\begin{equation}
    D_{ij}[\Phi,\Phi] = \Phi \partial_i \partial_j \Phi - \partial_i\Phi \partial_j \Phi - i A_{ij}\Phi^2 .
\end{equation}
Here $A_{ij}$ is a symmetric $U(1)$ tensor gauge field with the transformation properties  $A_{ij} \ra A_{ij} + \partial_i \partial_j \alpha$, while $\Phi \ra e^{i\alpha} \Phi$ under a gauge transformation. 

We would also need to add a kinetic term for this gauge field, and in general we one could investigate this tensor gauge theory in its own right. The gauge invariant field strength of this theory would be \cite{Seiberg:2020wsg}
\begin{equation}
    F_{[ij]k} = \partial_i A_{jk} - \partial_j A_{ik},
\end{equation}
and the kinetic term takes the form $F_{[ij]k} F^{[ij]k}$. We note that such a theory only makes sense in $d\geq 3$, similar to how standard gauge theory only makes sense in $d\geq 2$. From here one can continue studying conservation laws, defects, and dynamics, similar to the gauge theories considered in \cite{Seiberg:2020bhn,Seiberg:2020wsg,Pretko_2017,Pretko_2018}.

A different approach to systematically construct interacting theories from the classical XY-plaquette model is by promoting a global $U(1)$ symmetry to a subsytem symmetry. This procedure is reminiscent of the gauge principle, where one promote a global symmetry to a local one. 

For example we can start out with a field theory involving a single complex scalar field $\psi$ which is charged under a global $U(1)$ symmetry, and a Lagrangian
\begin{equation}
    L = \partial_\mu\psi^\dagger \partial^{\mu} \psi + V(\psi^\dagger \psi).
\end{equation}
Then we would like to partially gauge the global symmetry into a subsystem symmetry. This can be done by introducing a ``covariant" derivative
\begin{equation}
    D_\mu \psi = \left(\partial_\mu - ig \partial_\mu \phi \right)\psi,
\end{equation}
where $\phi$ acts as the analog of a gauge field. Under the subsystem symmetry $\phi$ transforms as \eqref{eq:symmetry_xy_classical}, while $\psi \ra e^{ig\sum_{\mu} f_\mu(x^\mu) }\psi$. The full action of this theory would then become
\begin{equation}
    L = D_\mu\psi^\dagger D^{\mu} \psi + V(\psi^\dagger \psi) + \sum_{\mu \neq \nu}\partial_{\mu} \partial_\nu \phi \partial^{\mu} \partial^\nu \phi .
\end{equation}
The resulting theory is an interacting field theory with a subsystem symmetry.

In essence one can partially gauge any theory with a conserved $U(1)$ charge, resulting in a whole new family of theories to play with. It would be interesting to investigate how these partially gauged theories behave, how to generalize them to non-Abelian symmetries, and perhaps to understand if they have analogous Coulomb and Higgs phases. Finally, we note that this partial gauging procedure can also be used to construct interacting theories coupled to the XY-plaquette models of \cite{Seiberg:2020bhn,Seiberg:2020wsg}, which would also be interesting to investigate.




\section*{Acknowledgments}
We would like thank Michael Hermele, Nathan Seiberg, Shu-Heng Shao and Willy Fischler for useful discussions. The work of AK and AR was supported, in part, by the U.S. Department of Energy under Grant No. DE-SC0022021 and a grant from the Simons Foundation (Grant 651440, AK). The work of JD was supported in part by the National Science Foundation under Grant No.~PHY--1914679.

\appendix

\section{Correlation functions in the Classical XY-plaquette model} \label{app:corr}

The structure of correlation functions of simple uncharged operators is very non-standard in models with subsystem symmetry. In particular these correlation functions tend to diverge even when the operators are far apart, but one of the spatial variables is close \cite{gorantla2021lowenergy}. Such UV divergences in IR correlation functions has been considered a hallmark of UV/IR mixing. As such, we are interested in computing the correlation functions of the simple operators $\partial_i\partial_j \phi$, and to observe any patterns of UV/IR mixing in the classical XY-plaquette model. We will focus our analysis on the theory in two and three dimensions for simplicity, and we think this also provides a general qualitative idea of the structure of the correlation functions even in higher dimensions.

\subsubsection*{2 dimensions:}

For $d=2$ we can compute the correlation function directly as
\begin{equation}
    \left<\partial_1\partial_2 \phi(x) \partial_1\partial_2(0) \phi \right>
     = g \int\frac{d^2 k}{(2\pi)^2} e^{i(k_1 x_1 + k_2 x_2)} = g \delta(x_1) \delta(x_2) .
\end{equation}

This correlation function is somewhat simplistic, but does contain some interesting features. Notice that the correlation function is indeterminate when $x_1$ or $x_2$ are zero, which resembles the UV/IR mixing of correlation functions in \cite{gorantla2021lowenergy}. However, unlike in the XY plaquette model of \cite{Seiberg:2020bhn,gorantla2021lowenergy}, this divergent behavior cannot be regularized by an IR regulator (say by putting the system on a torus,) but rather requires a UV regulator. A similar picture appears in 3 dimensions, though there the correlation functions have more structure.

\subsubsection*{3 dimensions:}
In three dimension we can compute the 2-point correlation function of $\partial_1\partial_2 \phi$ to be
\begin{equation}
\begin{aligned}
    \left<\partial_1\partial_2 \phi(x) \partial_1\partial_2(0) \phi \right>
     &= g \int\frac{d^3 k}{(2\pi)^3} \frac{k_1^2 k_2^2}{k_1^2k_2^2 + k_1^2k_3^2 + k_2^2k_3^2}e^{i(k_1 x_1 + k_2 x_2 + k_3 x_3)} \\
     &= \frac{g}{2} \int\frac{d^2 k}{(2\pi)^2} \frac{|k_1| |k_2|}{\sqrt{k_1^2+k_2^2} }e^{i(k_1 x_1 + k_2 x_2 ) - \frac{|k_1||k_2||x_3|}{\sqrt{k_1^2+k_2^2}}} \\
     & = \frac{g}{8\pi^2} \int d\theta dr r^2 |\sin\theta \cos \theta|
     e^{ir( \cos \theta x_1 + \sin \theta x_2 ) - r|\sin \theta \cos \theta||x_3|} \\
     &= \frac{g}{4\pi^2} \int_0^{2\pi}d\theta \frac{x_3\sin^2 \theta \cos^2 \theta\left(x_3^2\sin^2 \theta \cos^2 \theta - \left( x_1\cos\theta +  x_2\sin\theta \right)^2 \right)}{\left(x_3^2\sin^2\theta \cos^2 \theta +( x_1\cos\theta +  x_2\sin\theta)^2 \right)^3}
\end{aligned}
\end{equation}

This result is finite if $x_i \neq 0$ for $i=1,2,3$, but diverges when the two operators share one of the same coordinates. We can expand this integral around these limits to see how it diverges. If we take $x_3 \ra 0$ then we see that
\begin{equation}
    \left<\partial_1\partial_2 \phi(x_1,x_2 , x_3\ra 0) \partial_1\partial_2(0) \phi \right> 
    = \frac{g \sqrt{x_1^2 + x_2^2}}{8\pi |x_1 x_2|} \cdot x_3^{-2} + O(1).
\end{equation}
On the other hand if we take $x_1 \ra 0$ then
\begin{equation}
    \left<\partial_1\partial_2 \phi(x_1 \ra 0,x_2 , x_3) \partial_1\partial_2(0) \phi \right> 
    = \frac{g}{8\pi x_3^2} \cdot |x_1|^{-1} + O(1),
\end{equation}
and taking $x_2\ra 0$ results in a similar expression with $x_2$ replacing $x_1$. 

We can also work in cylindrical coordinates, defining $x_1 = \rho \cos(\delta)$, $x_2 = \rho \sin(\delta)$, in which case the 2-point function becomes
\begin{equation}
    \left<\partial_1\partial_2 \phi(x) \partial_1\partial_2(0) \phi \right> = `
      \frac{g}{4\pi^2 x_3^3} \int_0^{2\pi}d\theta \frac{\sin^2 \theta \cos^2 \theta\left(\sin^2 \theta \cos^2 \theta - \frac{\rho^2}{x_3^2}\cos^2(\theta - \delta) \right)}{\left(\sin^2\theta \cos^2 \theta + \frac{\rho^2}{x_3^2}\cos^2(\theta - \delta) \right)^3} ,
\end{equation}
where the integral is only a function of $\rho/x_3$. In the limit $x_3 \ll \rho$ we get
\begin{equation}
    \left<\partial_1\partial_2 \phi(x) \partial_1\partial_2(0) \phi \right>
     \approx \frac{g}{4\pi |\sin(2\delta)| x_3^2 \rho}, 
\end{equation}
while in the limit $ \rho \ll x_3$ the integral evaluates to 
\begin{equation}
    \left<\partial_1\partial_2 \phi(x) \partial_1\partial_2(0) \phi \right>
     \approx \frac{g |\sin \delta| + |\cos \delta|}{4\pi |\sin(2\delta)| x_3^2 \rho}, 
\end{equation}

In all these limits the 2-point function diverges even though the operators are separated. This is similar to the divergence of correlation functions as two operators are brought close together, only in this case the operators stay separated and only the separation in one of their coordinates is taken to be close. This behavior is similar to the one obsercved in \cite{gorantla2021lowenergy}.

However, unlike in \cite{Seiberg:2020bhn,gorantla2021lowenergy}, this divergence must be regulated by a UV regulator, and cannot be regulated by an IR regulator (i.e. putting the system in a finite volume.) To see this, we can take space to be a 3-Torus with equal lengths $\ell$, so that the momenta take the discrete values $k_i = 2\pi n_i/\ell$. Now the 2-point function becomes
\begin{equation}
    \begin{aligned}
    \left<\partial_1\partial_2 \phi(x) \partial_1\partial_2(0) \phi \right>
     &= \frac{g}{\ell^3} \sum_{n_1,n_2,n_3\in \mathbb{Z}} \frac{n_1^2 n_2^2}{n_1^2n_2^2 + n_1^2n_3^2 + n_2^2n_3^2}e^{i2\pi(n_1 x_1 + n_2 x_2 + n_3 x_3)/\ell} 
\end{aligned}
\end{equation}
Taking $x_1 = 0$ it is clear that the sum over $n_1$ at fixed $n_2,n_3$ does not converge, as is the case when taking $x_2=0$. If we take $x_3 = 0$ then we can still sum over $n_3$, giving us
\begin{equation}
    \begin{aligned}
    \left<\partial_1\partial_2 \phi(x_1,x_2,0) \partial_1\partial_2(0) \phi \right>
     &= \frac{g}{\ell^3} \sum_{n_1,n_2,n_3\in \mathbb{Z}} \frac{n_1^2 n_2^2}{n_1^2n_2^2 + n_1^2n_3^2 + n_2^2n_3^2}e^{i2\pi(n_1 x_1 + n_2 x_2 )/\ell}  \\
     &= \frac{\pi g}{\ell^3} \sum_{n_1,n_2\in \mathbb{Z}} \frac{|n_1 n_2|}{\sqrt{n_1^2 + n_2^2}} e^{i2\pi(n_1 x_1 + n_2 x_2 )/\ell} \coth\left(\frac{\pi|n_1||n_2|}{\sqrt{n_1^2+n_2^2}}\right),
\end{aligned}
\end{equation}
but now the sum over $n_1$ and $n_2$ is not convergent, and it still requires a UV regulator.

As this divergence cannot be regulated by IR regulator, it is unclear to us how these correlation functions fit into the picture of UV/IR mixing. Furthermore, one can define a re-normalized operator $\mathcal{O}_{12} \sim \Lambda^{3/2} \partial_x \partial_y \phi$, where $\Lambda$ is some UV scale. Then this renormalized operator would have a finite 2-point function of the form
\begin{equation}
    \left< \mathcal{O}_{12}(x_1,x_2,x_3) \mathcal{O}_{12}(y_1,y_2,y_3)\right> = \left\{ \begin{array}{cc}
    C,\quad & x_3-y_3 = x_1-y_1 = 0, \text{ or } x_3-y_3 = x_2-y_2 = 0, \\ 0, \quad &\text{otherwise,}\end{array} \right.
\end{equation}
where $C$ is some constant that depends on the re-normalization scheme. This renormalized operator has restricted mobility as it can only correlate along lines of fixed $x_1$ and $x_2$ in the $x_1$--$x_2$ plane, and in terms of these renormalized operators the theory would have no knowledge of the UV physics.

However we are unconvinced that this is the correct way to describe the IR theory, and are skeptical if this divergence of the 2-point function should be regulated at all. Rather, we think this divergence is contains important information about the IR physics, and the causal structure of the theory. An analogous situation arises in relativistic quantum field theory, where it is common for propagators diverge on the light cone. This light cone singularity is a genuine property of the IR physics and the structure of space-time, and not something which is needs to be regulated away. In any case it would be interesting to further understand the consequences of these divergences, and to fit them more squarely into the picture of UV/IR mixing.


\bibliographystyle{JHEP}
\bibliography{brokenfractons}

\end{document}